\documentclass[12pt,draftclsnofoot,journal,a4paper,onecolumn]{IEEEtran}

\usepackage{cite}
\usepackage{graphicx}
\usepackage{epstopdf}
\usepackage{amsmath}
\usepackage{amsfonts}
\usepackage{algorithmic}
\usepackage{array}
\usepackage{stfloats}
\usepackage{multirow}
\usepackage{color}

\begin{document}
\newcounter{MYtempeqncnt}

\title{Capacity Analysis of Decoupled Downlink and Uplink Access in 5G Heterogeneous Systems}

\author{Katerina Smiljkovikj, Hisham ElShaer, Petar Popovski,  Federico Boccardi, Mischa Dohler, Liljana Gavrilovska and Ralf Irmer
\thanks{K. Smiljkovikj and L. Gavrilovska are with the Faculty of Electrical Engineering and Information Technologies, Ss Cyril and Methodius University in Skopje, Macedonia, H. ElShaer, F. Boccardi and R. Irmer are with Vodafone R\&D UK, P. Popovski is with the Department of Electronic Systems
Aalborg University, Aalborg, Denmark and M. Dohler is with the Centre for Telecommunications Research at King's College London, London, UK.}
}


\maketitle

\begin{abstract}
Our traditional notion of a cell is changing dramatically given the increasing degree of heterogeneity in 4G and emerging 5G systems. Rather than belonging to a specific cell, a device would choose the most suitable connection from the plethora of connections available. In such a setting, given the transmission powers differ significantly between downlink (DL) and uplink (UL), a wireless device that sees multiple Base Stations (BSs) may access the infrastructure in a way that it receives the downlink (DL) traffic from one BS and sends uplink (UL) traffic through another BS. This situation is referred to as \emph{Downlink and Uplink Decoupling (DUDe)}.
In this paper, the capacity and throughput gains brought by decoupling are rigorously derived using stochastic geometry. Theoretical findings are then corroborated by means of simulation results. A further constituent of this paper is the verification of the theoretically derived results by means of a real-world system simulation platform. Despite theoretical assumptions differing from the very complete system simulator, the trends in the association probabilities and capacity gains are similar. Based on the promising results, we then outline architectural changes needed to facilitate the decoupling of DL and UL.
\end{abstract}


\IEEEpeerreviewmaketitle

\section{Introduction}

\IEEEPARstart{H}{eterogeneous} networks (HetNets) have been recognized as a, if not the, most promising approach to yield required communications rates in 4G and emerging 5G systems \cite{IsThePHYLayerDead,GC_2}. Heterogeneity here typically spans across Macro cells (Mcells) and different types of Small cells (Scells), such as Micro cells (Micell), Pico cells (Pcells) and Femto cells (Fcells); in the future, the inclusion of 3GPP-alien systems, such as WiFi, shall also be considered in this setting. However, the current 3GPP communications architecture and protocols were designed with Mcells in mind and heterogeneity was just an afterthought. Having heterogeneity in place, it is  important to present a fresh look on how the networks should be deployed and operated, as well as which fundamental improvements should be made in order to have efficient operation of the heterogeneous deployment model as a part of 5G wireless.

Notably, the increasing heterogeneity is dramatically changing our traditional notion of a communication cell \cite{GC_4}: as the number of Base Stations (BSs) becomes comparable to the number of devices and the deployment pattern of the BSs is rather irregular, there are multiple BSs from which a device can select one to associate with. Given that the uplink traffic is gaining in importance, transmission powers and interference levels differ significantly between available systems, and the link quality experience is very different between downlink (DL) and uplink (UL). The current policy for association to the BS is \emph{DL Reference signal received Power (DRP)} and works such that  the device is associated to a BS both for DL and UL transmissions, provided that the selected BS offers the highest downlink received power. One of the main contribution of this paper is to challenge the current policy used by a device to associate to a BS. Our prior work  \cite{DUDe,KaterinaPetarLiljanaDUD,GC_2,GC_3,GC_4} has commenced to explore the idea of completely decoupling the DL and the UL of a communications data flow. The exposed gains are significant and inspired further investigation presented in this work.

To understand this assertion we consider a typical HetNet scenario with a Mcell tier and a Scell tier, where in this paper we consider outdoor Scells only. The DL coverage of the Mcell is much larger than the Scell due to the large difference in their transmit powers. However, in the UL all the transmitters, which are battery-powered mobile devices, have the same transmit power and thus the same range. Therefore, a device that is connected to a Mcell in the DL from which it receives the highest signal level might want to connect to a Scell in the UL to which the path loss is the lowest. This above example is aggravated by the fact that HetNets become denser, the Scell BSs (SBSs) come closer to the users and thus the disparity between Mcells and Scells is increasing. As a consequence, the gap between the optimal DL and UL cell boundaries increases. This necessitates an entirely new design approach which we refer to as the \emph{Downlink and Uplink Decoupling (DUDe)}. Here, the UL and DL are treated as entirely separate network entities where the device is allowed to connect to different serving BSs in the UL and DL, respectively.

\subsection{Related Work}

The loosening and decoupling of the DL and UL association has been indicated in \cite{GC_2}, and has been studied in a few selected pioneering contributions \cite{DUDe,KaterinaPetarLiljanaDUD,GC_3,GC_4}.

Notably, the most recent contributions \cite{DUDe,KaterinaPetarLiljanaDUD} have studied the impact of having a cell association which is different for downlink and uplink. The authors of \cite{DUDe} have investigated the throughput and outage gains via the real-world simulation tool Atoll; the simulation results have yielded important performance gains which increase with an increasing density of small cells. The authors of \cite{KaterinaPetarLiljanaDUD} then studied the association probabilities in more details, i.e. the probabilities that a device would associate i) both DL and UL to the Mcell; ii) DL to Scell and UL to Mcell; iii) DL to Mcell and UL to Scell; and iv) both DL and UL to the Scell.

Earlier contributions \cite{GC_2,GC_3} introduced the early notion of DUDe. In particular, in \cite{GC_4}, DUDe is considered as a part of a broader ``device-centric'' architectural vision, where the set of network nodes providing connectivity to a given device and the functions of these nodes in a particular communication session are tailored to that specific device and session. A study in \cite{GC_5} tackles the problem from an energy efficiency perspective where the UL/DL decoupling allows for more flexibility in switching-off some BSs and also for saving energy at the terminal side.

One technique that brings some fairness to the UL is ``Range Extension'' (RE) where the idea is to add a cell selection offset to the reference signals of the Scells to increase their coverage in order to offload some traffic from the Mcells \cite{GC_7}. However, using offsets greater than 3-6 dB leads to high interference levels in the DL which is why techniques, such as enhanced Inter-Cell Interference Coordination (eICIC), have been developed to try to combat this type of interference \cite{GC_8}. Nevertheless, the RE technique is limited to moderate offset values due to the harsh interference in the DL. Here, the DUDe concept yields benefits of having very high RE offsets in the UL without the interference effects in the DL.

\subsection{Main Contributions}
The work in this paper extends prior art, most notably the recent work conducted in \cite{DUDe,KaterinaPetarLiljanaDUD}. The main contributions are summarized as follows:

\begin{itemize}

\item {\bf Analytical evaluation of the decoupling policy.} Based on stochastic geometry and prior derived association probabilities, we derive the achievable rates of a decoupled system. The derived expressions are easily evaluated numerically. Furthermore, corroborating simulations have been conducted.

\item {\bf Real-World Verification.} The decoupling methodology is applied to Vodafone's real-world system level simulation tool Atoll which adds a significant practical value to the findings.  Absolute real-world gains are obtained, as well as the trends which coincide with the theoretical findings.

\item {\bf Architectural Considerations.} We outline the important architectural changes needed to facilitate the decoupling, where we base our developments on a 3GPP baseline architecture.

\end{itemize}

It was very interesting to note that the analysis with stochastic geometry and real-world experimental data show the same trend in the association probabilities. This led us to verify the trend by performing additional simulations over a third deployment model, in which the BSs are placed in a regular grid. The trend was confirmed with the third model as well, thereby leading to a conclusion that the association probability depends chiefly on the deployment density and less on the actual deployment process.

%

The rest of the paper is organized as follows. In Section \ref{SystemModelSG}, we review the underlying system model based on stochastic geometry and present the analytical framework related to the capacity and spectral efficiency of the decoupled downlink/uplink system. In Section \ref{ExpSetting}, we describe in great details the real-world Atoll simulation framework and discuss the results of the simulations. Thereupon, in Section \ref{DiscussCompare} we discuss the similarities in the trends and we give general conclusions on decoupled access. The architectural changes incurred by the decoupling policy are discussed in Section \ref{architecture}. The paper is concluded with Section \ref{conclusions}.

\section{System model and analysis with stochastic geometry} \label{SystemModelSG}

In this section we analyze the decoupling of UL and DL using stochastic geometry model for heterogeneous networks, which has been recently used to gain theoretical insights in the performance of cellular networks \cite{ElSawyHaenggiSG}. We derive performance metrics in terms of probability of association in UL/DL and average capacity. Surprisingly, the trends in the results observed in the model with stochastic geometry are replicated by two substantially different deployment models, an experimental setting and a regular grid model.

\subsection{System model}

The system model represents heterogeneous cellular network, consisting of two tiers, Mcell tier and Scell tier. The locations of BSs and the locations of devices are modeled by independent homogeneous Poisson Point Processes (PPPs). Each PPP is denoted as $\Phi_v$ and has intensity measure $\lambda_v$, where $v=M$ for Mcells, $v=S$ for Scells and $v=d$ for devices. A point in $\mathbb{R}^2$ that is result of realization of $\Phi_v$ is denoted as $x_v=(x_{v_1},x_{v_2})$. The transmit power of Mcells, Scells and devices is $P_M$, $P_S$ and $P_d$, respectively. Without loss of generality, the analysis is performed on a typical device located at the origin, i.e. $x_d=(0,0)$. By Slivnyak's theorem, a PPP conditioned on a presence of a typical point in the origin has the same distribution as the original PPP \cite{KendallMeckeSGapps}.

We analyze the association probability (Section~\ref{AssocProb}), where we consider both directions, DL and UL, and we analyze the spectral efficiency (Section~\ref{SpectEff}), where we focus on the UL only. The signal power received from BS located at $x_v \in \Phi_v$ is denoted as $S_v^{DL}$ and the signal power received at the same BS in UL is denoted as $S_v^{UL}$. The signals are given by:
\begin{eqnarray}
S_v^{DL} &=& P_v h_{x_v} \chi_v \left\|x_v\right\|^{-\alpha} \label{DLsignal} \\
S_v^{UL} &=& P_d h_{x_v} \chi_v \left\|x_v\right\|^{-\alpha} \label{ULsignal}
\end{eqnarray}
\noindent
where $h_{x_v}$ describes Rayleigh fading and is an exponentially distributed random variable with unit mean. $\left\|x_v\right\|$ is the distance from $x_v$ to the origin and $\alpha$ is the path loss exponent ($\alpha>2$). $\chi_v$ is lognormal shadowing defined as $\chi_v = 10^{\frac{X_v}{10}}$, where $X_v \sim N(\mu_v,\sigma_v^2)$. We are using the approach elaborated in Lemma 1 in \cite{DhillonAndrews}, where the authors include the shadowing in a transparent way by using the displacement theorem \cite{BaccelliBlaszczyszyn}. The received signals given by~(\ref{DLsignal}) and~(\ref{ULsignal}) can be transformed in terms of the displaced points $y_v$:
\begin{eqnarray}
S_v^{DL}&=&P_v h_{x_v} \left\|\chi_v^{-1/\alpha} x_v\right\|^{-\alpha}=P_v h_{x_v} \left\|y_v\right\|^{-\alpha} \label{DLsignal_transformed} \\
S_v^{UL}&=&P_d h_{x_v} \left\|\chi_v^{-1/\alpha} x_v\right\|^{-\alpha}=P_d h_{x_v} \left\|y_v\right\|^{-\alpha} \label{ULsignal_transformed}
\end{eqnarray}
By displacement theorem, the points $y_v$ are obtained by independent realization of equivalent PPP $\widetilde{\Phi}_v$ with intensity $\widetilde{\lambda}_v$, which is related to the intensity of the original PPP by the fractional moment of $\chi_v$:
\begin{eqnarray}
\widetilde{\lambda}_v &=& \mathbb{E}\left[ \chi_v^{2/\alpha} \right]\lambda_v \nonumber \\
										  &=& \textrm{exp}\left( \frac{ln10}{5}\frac{\mu_v}{\alpha} + \frac{1}{2}\left(\frac{ln10}{5}\frac{\sigma_v}{\alpha}\right)^2 \right) \lambda_v	
\end{eqnarray}
In the remaining of the paper, we will use the equivalent processes $\widetilde{\Phi}_v$, where $v \in \{M,S,d\}$.


\subsection{Interference model}

Let $D_v$ denote the distance from the closest point from $\widetilde{\Phi}_v$ to the origin. The nearest point distance distribution of PPP is completely defined by the null probabilities of the process \cite{KendallMeckeSGapps}. The probability density function (pdf) and cumulative distribution	 function (cdf) are given by:
\begin{eqnarray}
	f_{D_v}(x) &=& 2\pi\widetilde{\lambda}_v x e^{-\pi\widetilde{\lambda}_v x^2}, x\geq0 \label{contactPDF} \\
	F_{D_v}(x) &=& 1-e^{-\pi\widetilde{\lambda}_v x^2}, x\geq0 \label{contactCDF}
\end{eqnarray}
We will use distributions in~(\ref{contactPDF}) and~(\ref{contactCDF}) to derive distance distributions to the serving BS in UL in Section~\ref{SpectEff}.

Recall that $\widetilde{\Phi}_d$ is PPP that describes the locations of devices. We assume that each BS avoids the interference among the devices associated to it through orthogonal resource allocation. Therefore, in UL the interference arises from devices associated to different BSs and transmit to the same resource unit. Having one interfering device from each BS, the number of interfering devices is equal to the number of BSs. Since the number of devices is larger than the number of BSs, then only a fraction of all devices $\widetilde{\Phi}_d$ causes UL interference. We are modeling the interfering devices by thinning the set $\widetilde{\Phi}_d$, sampling it randomly with probability $p=\frac{\widetilde{N}_{MS}}{\widetilde{N}_d}$, where $\widetilde{N}_{MS}=\widetilde{\lambda}_{M}A+\widetilde{\lambda}_{S}A$ is the average number of BSs in the area $A$ and $\widetilde{N}_{d}=\widetilde{\lambda}_dA$ is the average number of devices in the area. This representation is an approximation due to the dependence among the actual set of interfering devices created by the fact that each device needs to be associated to a different BS. However, as shown in \cite{UplinkModelingSG}, this dependence is weak and the random thinning is justified. The thinned process is denoted by $\widetilde{\Phi}_{I_d}$ and has an intensity of $\widetilde{\lambda}_{I_d}=p\widetilde{\lambda}_{d}$. Modeling interfering devices by thinning a point process is already used in \cite{KaterinaPetarLiljanaDUD}, where the authors show that it is as accurate as modeling all devices, associating them with BSs and randomly selecting one interferer per BS.

Using the notion of typical device located at the origin, one should calculate the Signal-to-Interference-plus-Noise Ratio (SINR) in the UL at a BS located at $y_v \in \widetilde{\Phi}_v$, which is not at the origin. This problem is simplified by translating the points from all point processes such that the associated BS in UL becomes located at the origin \cite{KaterinaPetarLiljanaDUD}. A homogeneous PPP is stationary, which means that the original and the translated versions have the same distribution for all points in $\mathbb{R}^2$ \cite{KendallMeckeSGapps}. Using the definition for signal power in~(\ref{ULsignal_transformed}), the UL SINR can be written as:
\begin{eqnarray}
SINR^{UL} = \frac{P_d h_{y_{v}} \left\|y_{v}\right\|^{-\alpha}}{\sum\limits_{{y_j} \in {\widetilde{\Phi}}_{I_d}} P_d h_{y_j} \left\|y_{j}\right\|^{-\alpha} + \sigma^2}
	\label{UL_SINR}
\end{eqnarray}
\noindent
where $\sigma^2$ is constant noise power at the receiver.

\subsection{Association probability}
\label{AssocProb}

The association decision is based on the average received signal in DL/UL, averaged over fading. By averaging~(\ref{DLsignal_transformed}) and~(\ref{ULsignal_transformed}), we obtain the signal powers that are used to decide the association:
\begin{eqnarray}
\mathbb{E}_{h}\left[ S_v^{DL} \right] &=& \mathbb{E}_{h} \left[ P_v h_{x_v} \left\|y_v\right\|^{-\alpha} \right] = P_v \left\|y_v\right\|^{-\alpha} \label{DLsignal_average} \\
\mathbb{E}_{h}\left[ S_v^{UL} \right] &=& \mathbb{E}_{h} \left[ P_d h_{x_v} \left\|y_v\right\|^{-\alpha} \right] = P_d \left\|y_v\right\|^{-\alpha} \label{ULsignal_average}
\end{eqnarray}

Following the policy that allows Downlink and Uplink Decoupling (DUDe), the device is associated as follows
\begin{align}
&\textrm{if }P_M D_M^{-\alpha}>P_S D_S^{-\alpha}\textrm{ then to a Mcell in DL} \label{eq:RuleDLMcell}\\
&\textrm{if }D_M^{-\alpha}>D_S^{-\alpha}\textrm{  then to a Mcell in UL} \label{eq:RuleULMcell}
\end{align}
Otherwise it is associated to a Scell. Combining both directions, there are four possible association cases. Each association case is defined by association region and association probability, derived in \cite{KaterinaPetarLiljanaDUD}. In this paper, we give only the final results, incorporating the fractional moment from displacement theory.

\subsubsection{Case 1: DL=UL=Mcell}
\begin{flalign}
		\Pr(\textrm{Case }1) &= \frac{\widetilde{\lambda}_M}{\widetilde{\lambda}_M + \widetilde{\lambda}_S}&&
		\label{Pcase1}
\end{flalign}

\subsubsection{Case 2; DL=Mcell \& UL=Scell}
\begin{flalign}
		\Pr (\textrm{Case }2) &= \frac{\widetilde{\lambda}_S}{{\widetilde{\lambda}_S} + {\widetilde{\lambda}_M}} - \frac{\widetilde{\lambda}_S}{{\widetilde{\lambda}_S} +  \left(\frac{P_M}{P_S}\right)^{2/\alpha} {\widetilde{\lambda}_M}}&&
		\label{Pcase2}
	\end{flalign}

\subsubsection{Case 3; DL=Scell \& UL=Mcell}
\begin{flalign}
	\Pr (\textrm{Case }3) &= 0&&
\end{flalign}

\subsubsection{Case 4; DL=UL=Scell}
\begin{flalign}
	\Pr (\textrm{Case }4) &= \frac{\widetilde{\lambda}_S}{\widetilde{\lambda}_S + \left(\frac{P_M}{P_S}\right)^{2/\alpha}\widetilde{\lambda}_M}&&
		\label{Pcase4}
\end{flalign}

We note that the conventional association policy, here referred to as DRP (DL Reference signal received Power), is implemented in the following way: If the condition (\ref{eq:RuleDLMcell}) is satisfied, then the device is associated to a Mcell in \emph{both} DL and UL; otherwise the device is associated to a Scell, again in both DL and UL.

\subsection{Spectral efficiency}
\label{SpectEff}

By decoupling UL and DL, we achieve improvement in the UL by adapting UL association to the actual conditions in the UL. Therefore, the analysis of spectral efficiency will be focused on UL only. Moreover, it is focused only on the devices with suboptimal association (devices that have closer Scell but receive higher downlink signal power from Mcell) because the other devices are not affected by path-loss based association in UL. The objective of this section is to analyze the spectral efficiency of the fraction of devices that have suboptimal association with DRP. Those are the devices that, using DRP, are associated to Mcell in both UL/DL, regardless of the fact that they have a closer Scell. Using DUDe, those devices are associated to a Scell in the UL and to a Mcell in the DL.

We derive the distribution of the distance to the serving BS in UL for both DRP and DUDe-based association. From \cite{KaterinaPetarLiljanaDUD}, the association region that corresponds to decoupled access is $\frac{P_S}{P_M}D_S^{-\alpha} < D_M^{-\alpha} \leq D_S^{-\alpha}$, where $P_S/P_M<1$. The distance to the serving BS is denoted as $D_{v,2}$, where $v=M$ with DRP and $v=S$ with DUDe. The second subscript, $2$, describes the conditioning on Case 2. When using DUDe, the complementary cumulative distribution function (ccdf) of the distance $D_{S,2}$ to the serving BS is derived as:
\begin{align}
		&F_{D_{S,2}}^c (x) = \Pr \left( D_S>x \mid \frac{P_S}{P_M}D_S^{-\alpha} < D_M^{-\alpha} \leq D_S^{-\alpha} \right) \nonumber \\
		&= \frac{\Pr \left( D_S > x ; D_S \leq D_M < \left(\frac{P_M}{P_S}\right)^{1/\alpha}D_S \right)}{\Pr(\textrm{Case }2)} \nonumber \\
		&= \frac{\int\limits_{x}^{\infty} \left(e^{-\pi \widetilde{\lambda}_M x_s^2} - e^{-\pi \widetilde{\lambda}_M \left(\frac{P_M}{P_S}\right)^{2/\alpha} x_s^2} \right) f_{D_S}(x_s) \mathrm{d}{x_s} } {\Pr(\textrm{Case }2)}
		\label{CondCDF}
	\end{align}
\noindent
where $\Pr(\textrm{Case }2)$ is given by~(\ref{Pcase2}). The cdf of the distance is $F_{D_{S,2}}(x) = 1 - F_{D_{S,2}}^c(x)$. By differentiating the cdf, we derive the pdf of the distance to the serving BS when DUDe is used, conditioned on Case 2:
\begin{eqnarray}
		f_{D_{S,2}} (x) &=& \frac{d F_{D_S}(x)}{dx} \nonumber \\
		&=& \frac{\left(e^{-\pi \widetilde{\lambda}_M x^2} - e^{-\pi \widetilde{\lambda}_M \left(\frac{P_M}{P_S}\right)^{\frac{2}{\alpha}} x^2} \right) f_{D_S}(x)} {\Pr(\textrm{Case }2)}
		\label{CondPDF_DUD}
	\end{eqnarray}
Following the same procedure, the distribution of the distance to the serving base station using DRP, conditioned on Case 2 is:
\begin{eqnarray}
		f_{D_{M,2}} (x) &=& \frac{\left(e^{-\pi \widetilde{\lambda}_S \left(\frac{P_S}{P_M}\right)^{\frac{2}{\alpha}} x^2} - e^{-\pi \widetilde{\lambda}_S x^2} \right) f_{D_M}(x)} {\Pr(\textrm{Case }2)}
		\label{CondPDF_NoDUD}
	\end{eqnarray}
When DUDe is used, the distance to the serving base station in~(\ref{UL_SINR}) has a pdf defined by~(\ref{CondPDF_DUD}). In the remaining part of this section we will derive the spectral efficiency for the decoupled access. The results for DRP association follow by substituting the distribution of the distance to the associated base station by~(\ref{CondPDF_NoDUD}).

The spectral efficiency, or equivalently, the normalized throughput with DUDe is defined as:
\begin{eqnarray}
		C_{DUDe} = \mathbb{E}\left[\textrm{log}_2 \left( 1 + SINR^{UL} \right) \right]
 		\label{spectralEff}
\end{eqnarray}
For $T>0$, $\mathbb{E}[T]=\int\limits_0^\infty \Pr (T>t) \mathrm{d}t$. Applying this property in equation~(\ref{spectralEff}), the spectral efficiency reduces to:
\begin{eqnarray}
		C_{DUDe} &=& \frac{1}{\textrm{ln}(2)} \int\limits_0^\infty \Pr \left( \textrm{ln} \left( 1 + SINR^{UL}\right) > t \right) \mathrm{d}{t} \nonumber \\
		&=& \frac{1}{\textrm{ln}(2)} \int\limits_0^\infty \Pr \left( SINR^{UL} > e^t-1 \right) \mathrm{d}{t}
 		\label{spectralEff2}
\end{eqnarray}
The integrand in~(\ref{spectralEff2}) is basically a definition for coverage probability, with SINR threshold set to $e^t-1$. It is evaluated as:
\begin{align}
	&\Pr \left( SINR^{UL} > e^t-1 \right) = \nonumber \\
	&= \Pr \left(\frac{P_d h_{y_{S}} D_{S,2}^{-\alpha}}{I_y + \sigma^2} > e^t-1\right) \nonumber \\
	&= \mathbb{E}_y\left[P \left( h_{y_{S}} > (e^t-1) y^{\alpha} \left(I_{y} + \sigma^2\right) | D_{S,2}=y \right) \right]= \nonumber \\
	&= \int\limits_{0}^{\infty} {\mathbb{E}_{I_y} \left[ e^{-(e^t-1) y^\alpha I_y}\right] e^{-(e^t-1) y^\alpha \sigma^2} f_{D_{S},2}(y)} \mathrm{d}{y} = \nonumber \\
	&= \int\limits_{0}^{\infty} {L_{I_y} \left( (e^t-1) y^\alpha \right) e^{-(e^t-1) y^\alpha \sigma^2} f_{D_{S},2}(y)} \mathrm{d}{y}
\label{P_coverage}
\end{align}
where $L_{I_y} \left( (e^t-1) y^\alpha \right)$ is the Laplace Functional (LF) of the interference, derived as
\begin{align}
	&L_{I_y} \left( s \right) = \mathbb{E}_{I_y} \left[ e^{-s I_y}\right] = \nonumber \\
		&= E_{\widetilde{\Phi}_{I_d}} \left[\prod\limits_{{y_j} \in \widetilde{\Phi}_{I_d}} \mathbb{E}_h \left[ e^{-s h_{y_j} \left\|y_{j}\right\|^{-\alpha}} \right] \right] \nonumber \\
		&= \exp \left( -2\pi \widetilde{\lambda}_{I_d} \int\limits_0^\infty \left(  1-\frac{1}{1+s v^{-\alpha}} \right) v \mathrm{d}{v} \right)
	\label{Pc_LF}  	
\end{align}
Combining~(\ref{CondPDF_DUD}),~(\ref{P_coverage}) and~(\ref{Pc_LF}), we derive the final expression for spectral efficiency with decoupled access conditioned on devices with suboptimal association (Case 2), given in equation~(\ref{C_DUD_final}). The expression for spectral efficiency with DRP for devices with suboptimal association has a similar form to (\ref{C_DUD_final}) and is given by (\ref{C_NoDUD_final}).

\begin{figure*}[!t]
\normalsize
\setcounter{MYtempeqncnt}{\value{equation}}
\setcounter{equation}{23}

\begin{eqnarray}
C_{DUDe} = \frac{\textrm{log}_2(e)}{\Pr(\textrm{Case }2)} \int\limits_0^\infty \int\limits_0^\infty {e^{-\pi \widetilde{\lambda}_{I_d} (e^t-1)^{\frac{2}{\alpha}} y^2 \int\limits_0^\infty \left(\frac{1}{1+ v^{\alpha/2}} \right) \mathrm{d}{v}} e^{-\frac{(e^t-1) y^\alpha \sigma^2}{P_d}}} \times \nonumber \\
\left(e^{-\pi \widetilde{\lambda}_M y^2} - e^{-\pi \widetilde{\lambda}_M \left(\frac{P_M}{P_S}\right)^{\frac{2}{\alpha}} y^2} \right) 2\pi\widetilde{\lambda}_S y e^{-\widetilde{\lambda}_S \pi y^2} \mathrm{d}{t} \mathrm{d}{y}
\label{C_DUD_final}
\end{eqnarray}

\begin{eqnarray}
C_{DRP} = \frac{\textrm{log}_2(e)}{\Pr(\textrm{Case }2)} \int\limits_0^\infty \int\limits_0^\infty {e^{-\pi \widetilde{\lambda}_{I_d} (e^t-1)^{\frac{2}{\alpha}} y^2 \int\limits_0^\infty \left(\frac{1}{1+ v^{\alpha/2}} \right) \mathrm{d}{v}} e^{-\frac{(e^t-1) y^\alpha \sigma^2}{P_d}}} \times \nonumber \\
\left(e^{-\pi \widetilde{\lambda}_S \left(\frac{P_S}{P_M}\right)^{\frac{2}{\alpha}} y^2} - e^{-\pi \widetilde{\lambda}_S y^2} \right) 2\pi\widetilde{\lambda}_M y e^{-\widetilde{\lambda}_M \pi y^2} \mathrm{d}{t} \mathrm{d}{y}
\label{C_NoDUD_final}
\end{eqnarray}

\setcounter{equation}{\value{MYtempeqncnt}}

\hrulefill
\vspace*{4pt}
\end{figure*}

\subsection{Numerical results}

\begin{figure}[h!]
	\centering
		\includegraphics[width=0.8\textwidth]{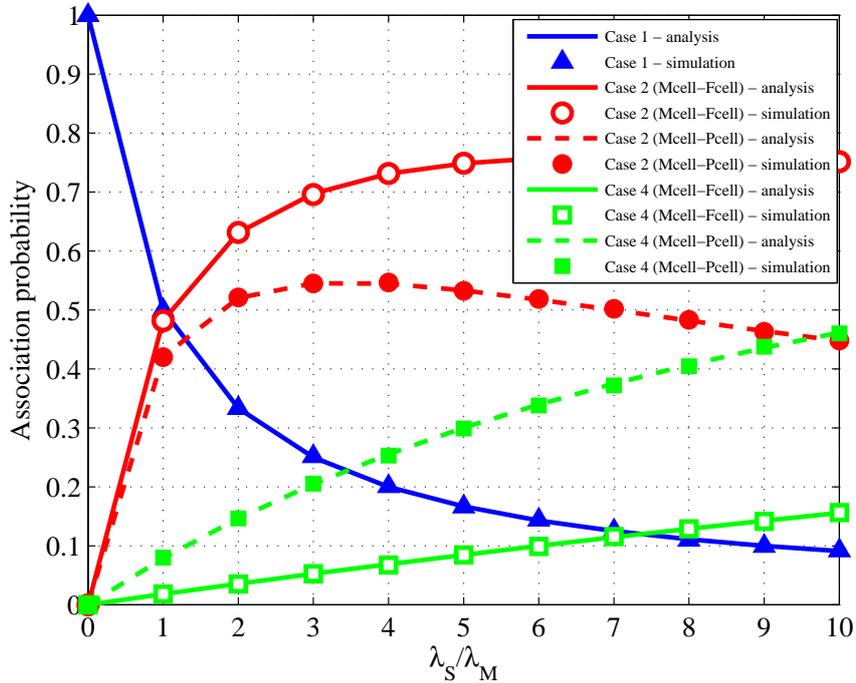}
		\caption{Association probability for Mcell-Fcell and Mcell-Pcell heterogeneous network ($P_M=46$ dBm, $P_S=20$ dBm for Mcell-Fcell; $P_S=30$ dBm for Mcell-Pcell, $P_d=20$ dBm, $\alpha=3$).}
		\label{fig:AssocProb}
\end{figure}

The analysis presented in previous sections is validated by numerical results. The association probabilities for joint association in DL and UL are analyzed for two types of Scells, high power Scells - Pcells and low power Scells - Fcells; here the high/low power refers to the power used by the base station of the particular Scell. Fig.~\ref{fig:AssocProb} shows association probabilities for a Mcell-Fcell and Mcell-Pcell two-tier network. In both cases, it is visible that increasing the number of Scells rapidly increases the probability for decoupled access, which also corresponds to the percentage of devices with decoupled access. It is important to notice that for the Mcell-Fcell network, this percentage goes nearly to $75\%$. This corroborates that a high percentage of devices in today's networks have suboptimal association and thus achieve suboptimal performance. For Mcell-Pcell networks, the percentage of devices with decoupled access is slightly over $50\%$ because high power Scells force the devices to connect to Scells in DL, thus increasing the probability for Case 4 (DL/UL access with Scells).

By further increase of Scells density, the probability for decoupled access decreases at the expense of increased probability for Case 4. Basically, the difference between the power levels of the two tiers reflects in the association process, making trade-off only between the devices with decoupled access and devices with DL/UL association to Scells. The higher the disparity between $P_M$ and $P_S$, the higher the probability of decoupled access, as follows from the association rules (\ref{eq:RuleDLMcell}) and (\ref{eq:RuleULMcell}).
The percentage of devices that associate to Mcells in both directions constantly decreases.

\begin{figure}[h!]
	\centering
		\includegraphics[width=0.8\textwidth]{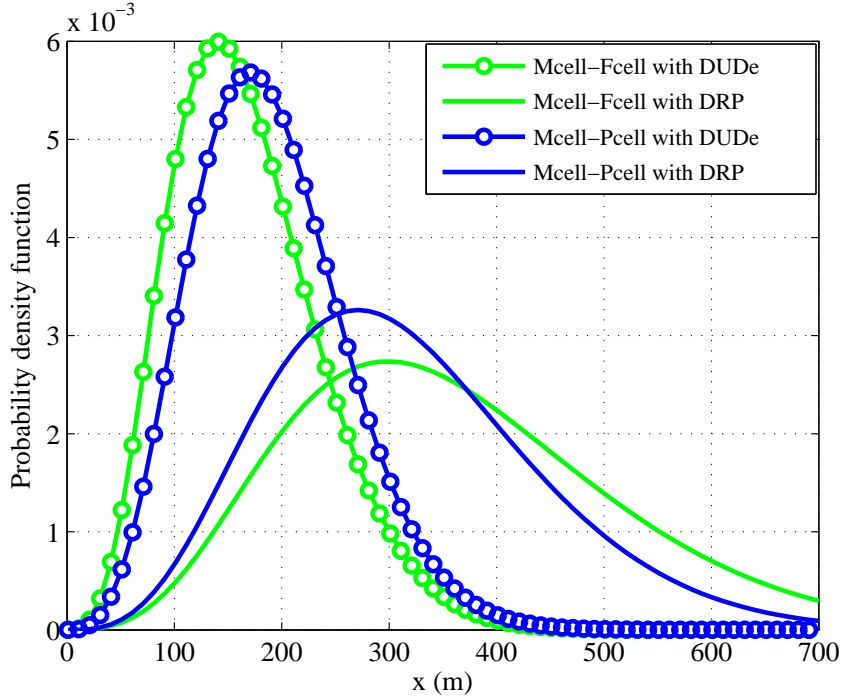}
		\caption{Probability density function of the distance to the serving BS ($P_M=46dBm, P_S=20dBm$ for Mcell-Fcell; $P_S=30dBm$ for Mcell-Pcell, $\lambda_S=5\lambda_M, \alpha=4$).}
		\label{fig:DistanceDist}
\end{figure}

The distance distributions to the serving BS with DUDe and with DRP, derived in Section~\ref{SpectEff}, are shown in Fig.~\ref{fig:DistanceDist}. Both network settings are included, Mcell-Fcell and Mcell-Pcell, respectively. By decoupling DL/UL, the distance distribution becomes narrower and shifts on the left towards smaller distances, i.e. there is higher probability that the serving BS will be closer to the device. This conclusion holds for both Fcells and Pcells. For the same BS density the pdf of the distance to the serving BS with DUDe for Mcell-Fcell network is narrower and is shifted to the left compared to the same distribution for Mcell-Pcell network. This is due to the fact that Pcells, with their higher transmit power, are able to associate more devices than Fcells and hence only the farthest devices remain in Case 2. On the other hand, Fcells have smaller coverage and therefore, there are many devices that are close to the Fcell but remain in Case 2. This percentage of devices increases the probability that the serving BS is closer to the devices that belong to Case 2 for Mcell-Fcell network.

\begin{figure}[h!]
	\centering
		\includegraphics[width=0.8\textwidth]{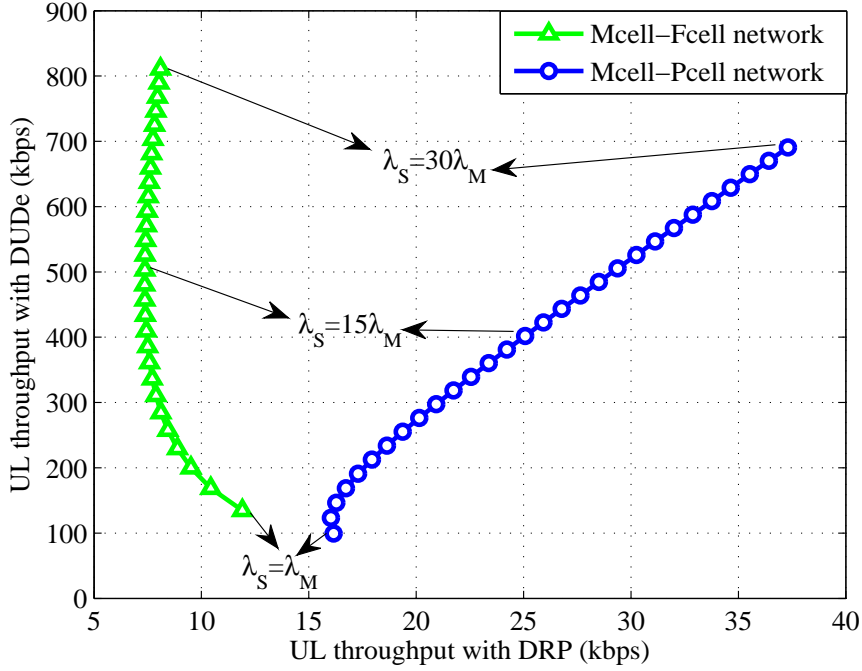}
		\caption{Spectral efficiency for Case 2 devices with and without decoupled access ($P_d=20dBm, \alpha=4, \lambda_d=10^4/km^2, B=20MHz$).}
		\label{fig:SE}
\end{figure}

By using DUDe, one can achieve a significant improvement in UL signal power at the associated BS. This results in increased SINR and thus improved spectral efficiency in the system. The results in Fig.~\ref{fig:SE} compare the UL throughput with DUDe and DRP for both network settings. The UL throughput with DUDe is calculated by multiplying the spectral efficiency by $B/N_a$, where $B$ is the system frequency bandwidth and $N_a$ is the average number of associated devices per BSs. Given the average number of devices in the area $N_d$ and the average number of BSs $N_{MS}=N_M+N_S$, $N_a=N_d/N_{MF}$. When DRP association is applied, $N_a$ is calculated using the association probabilities, i.e. $N_a=N_d(\Pr(\textrm{Case }1)+\Pr(\textrm{Case }2))/N_M$.

The abscissa shows the UL throughput with DRP and the ordinate shows the value with decoupled access for same BS density. It can be noted that the gains are significant. For instance,
for Mcell-Fcell network, the UL throughput of $10$ kbps with DRP maps into $>150$ kbps with DUDe. For the Mcell-Pcell network, the UL throughput of $30$ kbps with DRP maps into $500$ kbps with DUDe.

Another interesting observation is that Mcell-Pcell network achieves higher throughput with DRP  than Mcell-Fcell network, for the same BS density. On the other hand, Mcell-Fcell is superior in UL throughput with DUDe. If we observe the point for $\lambda_S=15\lambda_M$, we can see that with DRP, Mcell-Pcell achieves $~25$ kbps, while Mcell-Fcell network achieves $~7$ kbps. Using DUDe leads to the opposite situation, Mcell-Fcell achieves $~500$ kbps and Mcell-Pcell  achieves $~400$ kbps. This phenomenon is a consequence of the distance distributions presented in Fig.~\ref{fig:DistanceDist}. Basically, using high power Scells (Pcells), we force the devices to associate to Scells in both directions, leaving the worst case devices with DUDe. When using low power Scells (Fcells), we have additional percentage of devices in Case 2, which are closer to Scells and contribute to higher UL throughput with DUDe.

\section{Experimental evaluation of decoupled access} \label{ExpSetting}

In this section we study the decoupled access under a different system model, based on real world experimental setting. We use the same association rules as in the model with stochastic geometry. However, the randomized deployment rooted in experimental data is fundamentally different from the randomized deployment based on stochastic geometry and that is why it is rather surprising that one can observe the same trends in the association probabilities.

\subsection{Simulation setup}

For our simulation-based evaluation, we use a multi-technology radio planning tool called Atoll \cite{Atoll} in conjunction with a high resolution 3D ray tracing path loss prediction model \cite{ray_tracing}. This model takes into account clutter, terrain and building data which in turn guarantees a realistic and accurate modeling of the propagation conditions. Atoll allows performing system level simulations representing a snapshot of an LTE network. For each simulation, Atoll generates a device distribution using a Monte Carlo algorithm. The device distribution is based on realistic traffic data extracted from a live network. Resource allocation in each simulation is carried out over a period of 1 second (i.e. 100 LTE frames).
We use fractional UL power control as specified in \cite{power_control}. 
As a deployment setup, we use a Vodafone LTE small cell testbed network that is currently running in the London area. The test network covers an area of approximately one square kilometer. We use this existing testbed in order to simulate a relatively dense HetNet scenario.

\begin{figure}[h!]
	\centering
		\includegraphics[width=8cm]{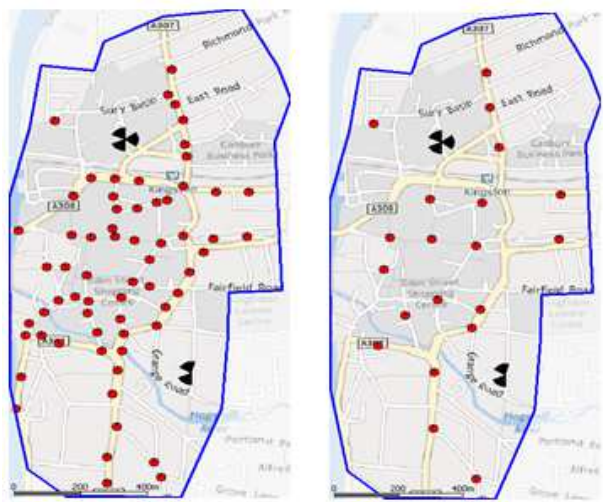}
		\caption{Fcells deployment (Left). Pcells deployment (Right).}
		\label{fig:sim_setup}
\end{figure}

In our simulations we consider two deployment scenarios. The first scenario is a dense deployment of Fcells where the transmit power of the Fcells is in the order of 20 dBm and the average coverage area per node is 20 meters square. The second scenario is a less dense deployment of Pcells where the transmit power of the Pcells is in the order of 30 dBm and the average coverage area per node is about 60 meters square. The two scenarios are illustrated in Fig. \ref{fig:sim_setup} where the black shapes and red circles represent the Mcell BSs and Scell BSs  respectively.
We consider a realistic device distribution based on traffic data from the field trial network during peak times. The distribution is up-scaled to simulate a high user density.

We compare the proposed DUDe mechanisms, where UL cell association is based on path loss, with the LTE baseline \cite{LTE_ref}, where cell association is based on DRP (DL Reference signal received Power ). DRP and DUDe association schemes are represented by~(\ref{eq:RuleDLMcell}) and~(\ref{eq:RuleULMcell}) respectively in the theoretical part in Section~\ref{SystemModelSG}.
In the case of DL with DRP we simulate two cases, Fcells and Pcells, respectively, with different deployment densities as discussed before. As in the case with stochastic geometry, this approach helps to understand the gains of DUDe in different deployment scenarios and different Scell transmit powers.
All the results in the next section will focus on the UL performance. The simulation parameters are listed in Table \ref{tab:sim_parameters}, where we consider an outdoor LTE network deployment.
Finally, we assume an ideal mechanism for delivering UL related DL signaling, such as acknowledgments, scheduling grants, etc. The reader is referred to Section~\ref{architecture} for more details about architectural aspects.



\begin{table}[!t]
\caption{Simulation parameters}
\centering
\begin{tabular}{|c||c|}
\hline
Operating frequency & 2.6 GHz (co-channel deployment)\\
\hline
Duplex mode & FDD\\
\hline
Bandwidth & 20 MHz (100 frequency blocks)\\
\hline
\multirow{3}{*}{Network deployment} & 	5 Mcells	\\
                       & 64 Fcells (case 1)\\
                       & 20 Pcells (case 2)
\\
\hline
\multirow{2}{*}{Device distribution} & 530 devices distributed according to traffic\\
                                    &  maps extracted from a live network
\\
\hline
Scheduler & Proportional fair\\
\hline
Simulation duration & 50 simulation runs (1 second each).\\
\hline
Traffic model & Full buffer\\
\hline
Propagation model & 3D ray-tracing model\\
\hline
\multirow{4}{*}{Max. transmit power} & 	Mcells = 46 dBm\\
                      &	Pcells = 30dBm\\
                      &	Fcells = 20 dBm\\
                      &	devices = 20 dBm
\\
\hline
\multirow{3}{*}{Antenna system} & 	Mcell: 2Tx, 2Rx, 17.8 dBi gain\\
                 &	Scell: 2Tx, 2Rx, 4 dBi gain\\
                 &	device: 1Tx, 1Rx, 0 dBi gain
\\
\hline
devices mobility & Pedestrian (3km/h)\\
\hline
UL modulation schemes & QPSK, 16 QAM, 64 QAM\\
\hline
\end{tabular}
\end{table}
\label{tab:sim_parameters}

\subsection{Simulation results} \label{results}

In this section, we present results comparing three scenarios:

\begin{itemize}
	\item
	DRP-based cell association where we have Fcells with low transmit power of $20$ dBm and the deployment shown in Fig.~\ref{fig:sim_setup} (Left). This case is referred to as Mcell-Fcell with DRP.
	\item
	DRP-based cell association where we have Pcells with high transmit power of $30$ dBm and the deployment shown in Fig.~\ref{fig:sim_setup} (Right). This case is referred to as Mcell-Pcell with DRP.
	\item
	
DUDe is represented by a cell association based on the path loss -- the transmit power of Scells  is irrelevant as the UL cell association is not based on DRP. The results for DUDe are produced for the same two deployment scenarios, Mcell-Fcell and Mcell-Pcell.
\end{itemize}

In our simulations we define a minimum and maximum throughput demand per device. A device has to reach the minimum throughput requirement to be able to transmit its data; otherwise, it is considered to be in outage. On the other hand, the maximum throughput demand puts a limit to the amount of throughput that each device can get, such that setting a high value for it helps in simulating a highly loaded network. For the next results we set the minimum and maximum throughput values to $200$ kbps and $20$ Mbps, respectively.

The scheduler applied in the system at first tries to satisfy the minimum throughput requirements for all the devices and, as a second step, it distributes the remaining resources among the devices to satisfy the maximum throughput demand of each device according to the proportional fair criterion. We note that the throughput results in the experimental setting are measured by using an actual scheduler and sectored Mcells, which makes the setting very different from the one with stochastic geometry and the throughput results are not comparable.

\begin{figure}[t!]
	\centering
		\includegraphics[width=0.8\textwidth]{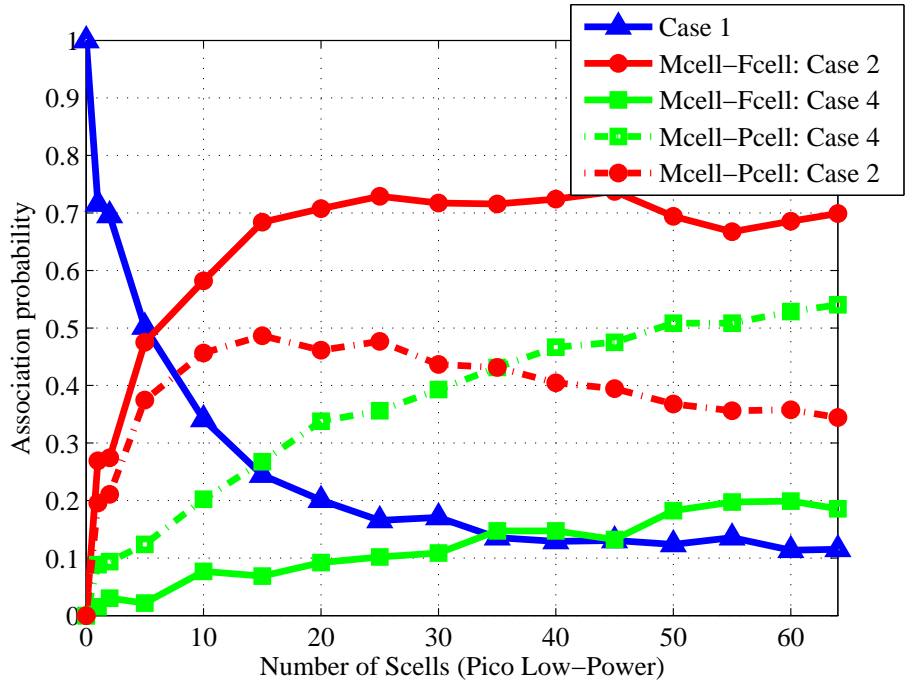}
		\caption{Association probability for the different association cases.}
		\label{fig:association}
\end{figure}

On the other hand, surprisingly, the different deployment setup does not affect the trend in the association probabilities, as shown on Fig. \ref{fig:association}. This figure shows device association probability as a function of the number of Scells for the different cases discussed in Section~\ref{AssocProb}. In this result we assume that the Mcell-Pcell case has $64$ Pcells which allows us to study how the cell association evolves with a high number of cells.
We note that Case 3 is ignored here since the association probability is always 0. Case 1 is the same for Mcell-Fcell and Mcell-Pcell and Cases 2 and 4 are plotted for the Mcell-Fcell and Mcell-Pcell cases.

The figure confirms the same trends shown in the analysis in Section~\ref{SystemModelSG}. Considering Mcell-Fcell, Case 2 is dominating where the probability of association saturates at around $70$\% starting from 20 Scells. Case 4 however is increasing with a very slow rate which is a result of the very limited DL coverage of the Fcells.
Looking at Mcell-Pcell, Case 2 is dominating with a probability of $50$\% up to a certain number of Scells (around $35$) where Case 4 surpasses Case 2. We notice that for Case 4 the curve is increasing with a higher slope than for Mcell-Fcell which can be explained by the fact that Pcells have a larger coverage area, such that after a certain density of Pcells most of the devices are connected to the Pcells in the UL and DL.
Another point to note in this figure is the comparison between the DL coverage of Pcells and Fcells, which is shown by comparing Case 4 for Mcell-Fcell and Mcell-Pcell, where we see that $10$ Pcells deliver the same level of coverage as $50$ Fcells. This comparison also shows that the probability of association in UL and DL to a Pcell is more than twice that to a Fcell with $64$ Scells deployed.

\begin{figure}[t!]
	\centering
		\includegraphics[width=0.8\textwidth]{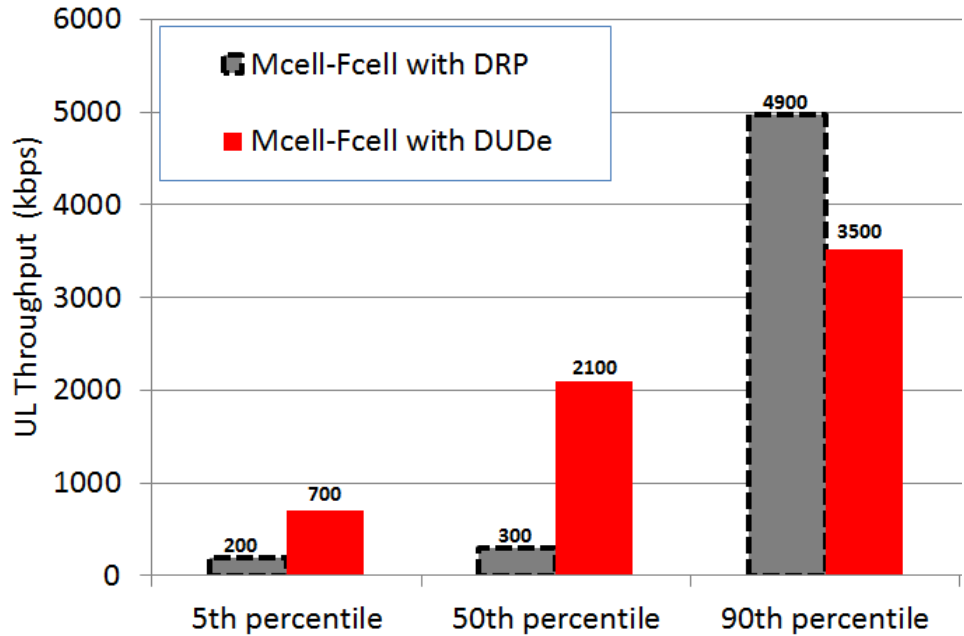}
		\caption{5th, 50th and 90th percentile UL throughput results for Mcell-Fcell network with and without DUDe.}
		\label{fig:femto_percent}
\end{figure}

\begin{figure}[t!]
	\centering
		\includegraphics[width=0.8\textwidth]{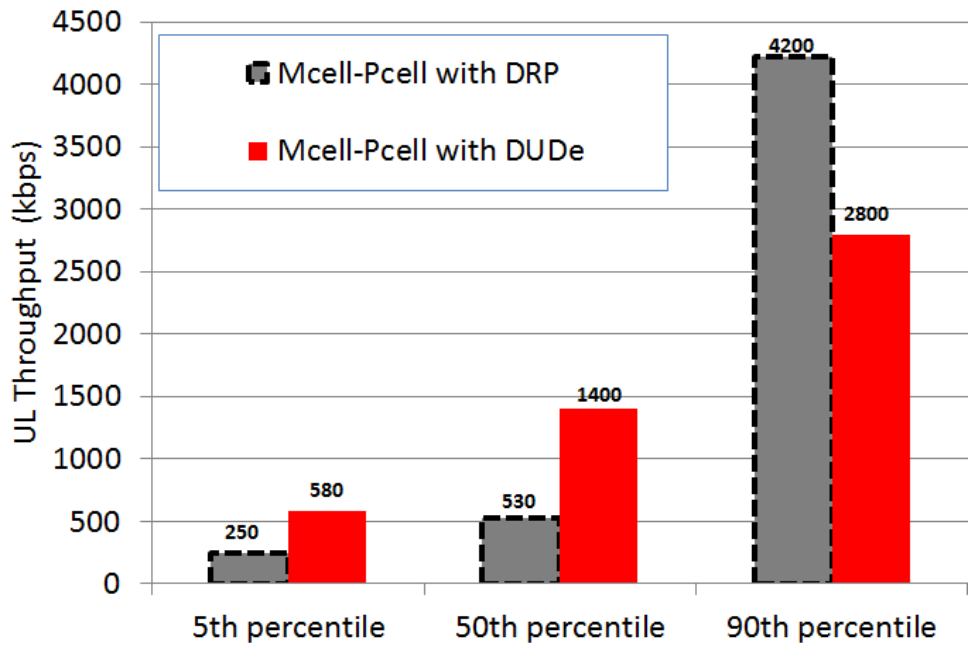}
		\caption{5th, 50th and 90th percentile UL throughput results for Mcell-Pcell network with and without DUDe.}
		\label{fig:pico_percent}
\end{figure}

Fig. \ref{fig:femto_percent} shows the comparison of the $5$th, $50$th, and $90$th percentile UL device throughput for Mcell-Fcell network with and without DUDe, whereas Fig. \ref{fig:pico_percent} shows the same type of comparison for Mcell-Pcell network.

In Fig. \ref{fig:femto_percent} we see that the $5$th and $50$th percentile throughput in the DUDe case are improved by more than $200$\% and $600$\% respectively. The gains in the 5th and 50th percentile are resulting from the higher coverage of the Scells in the DUDe case which results in a better distribution of the devices among the nodes and a much more efficient usage of the resources.
Also the fact that the devices connect to the node to which they have the lowest pathloss helps in reducing the UL interference as shown in \cite{DUDe}. This results in a higher device SINR that allows the devices to use a higher modulation scheme and in turn achieve a better utilization of the resources and a higher throughput.

Looking at the $90$th percentile throughput, we see that Mcell-Fcell with DRP association delivers a higher throughput than DUDe. This is expected, since Fcells have smaller DL coverage, such that they serve a much lower number of devices in the UL than DUDe and therefore there is an increase in the $90$th percentile throughput. However, this occurs at the expense of the $5$th and $50$th percentile throughput.

Fig. \ref{fig:pico_percent} shows the same behavior as Fig. \ref{fig:femto_percent} where DUDe results in an improvement in the $5$th and $50$th percentile throughput over the case with DRP by over $100$\% and $160$\% respectively.
Mcell-Pcell with DRP has a better $90$th percentile throughput which is a result of the lower number of devices that the Pcells serve, as discussed before.

Comparing Mcell-Fcell and Mcell-Pcell with DRP in figures \ref{fig:femto_percent} and \ref{fig:pico_percent}, we see that Mcell-Pcell provides a higher $5$th and $50$th percentile throughput whereas Mcell-Fcell provides a higher $90$th percentile throughput. However, due to the much lower coverage of Fcells in Mcell-Fcell, one would expect that the $90$th percentile throughput is much higher compared to Mcell-Pcell. This can be explained by looking at the $98$th percentile of throughput, where we see that Mcell-Pcell is about $10$ Mbps whereas Mcell-Fcell is $16.5$ Mbps, which is over $50$\% increase in throughput over Mcell-Pcell. This shows that the effect of Fcells in Mcell-Fcell is more visible on a limited number of devices.
These results shows the different use cases for Scells deployments, where Fcells would mainly be deployed to provide higher capacity to a low number of devices and Pcells to provide both capacity and coverage.
Furthermore, comparing the two network settings with DUDe in figure 5 and 6 we see that adding more Scells, i.e. comparing Fig. \ref{fig:femto_percent} where we have 64 Fcells to Fig. \ref{fig:pico_percent} where we have 20 Pcells, does not contribute much to the $5$th percentile throughput where the contribution is more in the $50$th and $90$th percentiles. This happens due to the fact that the $5$th percentile devices become interference-limited after a certain point.

The Range Extension (RE) technique basically works in the same direction as decoupling the UL and DL in a sense that it results in an increased coverage in the UL \cite{range_extension}. The disadvantage of RE is that the interference level in the DL increases aggressively as the RE bias increases, which requires the usage of interference management techniques \cite{icic}. This is not required with DUDe since UL and DL are treated as two different networks in this technique.

\begin{figure}[t!]
	\centering
		\includegraphics[width=0.8\textwidth]{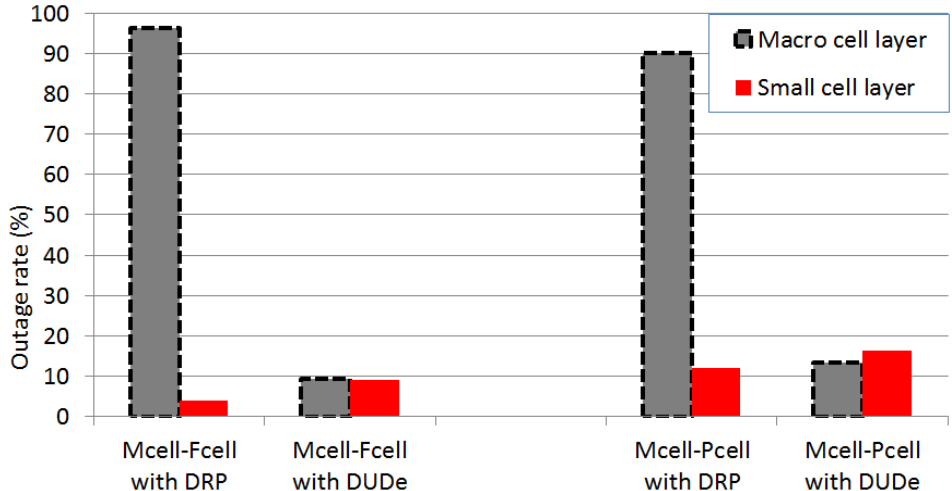}
		\caption{Outage probability comparison for Mcell-Fcell and Mcell-Pcell network settings, with and without DUDe.}
		\label{fig:outage}
\end{figure}

In the next result we run the same simulation, but increasing the minimum throughput demand to $1$ Mbps to study the outage rate in this high density scenario with a very high traffic demand.
Fig.~\ref{fig:outage} shows the average outage rate for the Mcell layer and the Scell layer where we compare both network settings, Mcell-Fcell and Mcell-Pcell, with DUDe and DRP, respectively.  The outage rate is defined as the percentage of the devices that fail to achieve the minimum throughput demand ($1$ Mbps) out of the total number of connected devices to a certain node. Since the simulated scenario is considered to be a highly dense one, it requires a very efficient use of resources in order to satisfy the high requirements of the devices. As seen in the figure, the Macro layer has a very high outage rate ($>90$\%) in the cases with DRP, which is basically explained by the fact that the Macro layer is very congested in the UL. Mcells do not have enough resources to serve all their devices at a high throughput level. However, when DUDe is applied, the devices are distributed more evenly between the nodes so the outage rate in that case is low (around $10-15$\%) for both Macro and Small cell layers.

These results clearly show that decoupling UL and DL where UL is based on path loss is a promising candidate for future networks, in which the network load is expected to increase in the UL and the priority is to provide a consistent where providing a consistent and ubiquitous service to all devices in different network deployments.

\section{Discussion and comparison}
\label{DiscussCompare}

In the previous sections, we have elaborated the concept of decoupled access with two particularly different settings, one based on theoretical model and one based on experimental data. Although both models have different assumptions and parameters, the trends in association probabilities are observable in both settings. This leads us to the conjecture that \emph{the association probability depends chiefly on the density of the deployment, but not the process used to generate the deployment geometry}.

\begin{figure}[h!]
	\centering
		\includegraphics[width=0.8\textwidth]{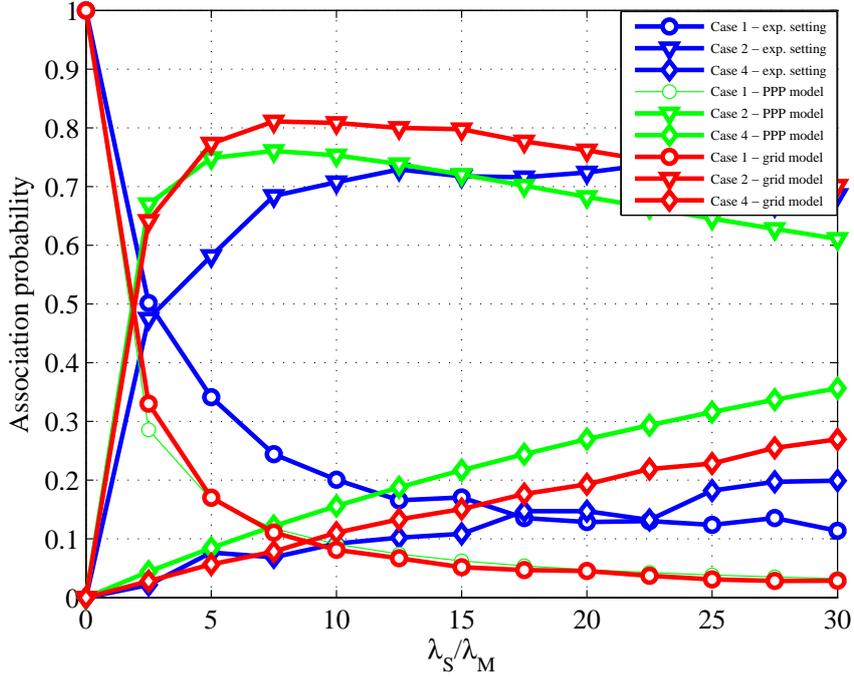}
		\caption{Association probability for three different network settings: experimental model, PPP model and grid model.}
		\label{fig:AssocProbCompare}
\end{figure}

In order to test this conjecture, we have also evaluated the association probabilities in a \emph{third} scenario, in which the BSs are deployed in a regular grid. The grid model is generated in the following way: $N_{MS}$ BSs are positioned in a grid such that they cover the same area $A$ as the BSs in the stochastic geometry model; each of them is assigned with transmit power $P_M$ with probability $Q$ and with transmit power $P_S$ with probability $B=1-Q$. Using the values of $Q$ and $B$ we can manipulate with the densities of the Mcells and Scells. The results for the association probability are shown in Fig.~\ref{fig:AssocProbCompare} and it is clearly visible that they are favorable to our conjecture. Using grid model to prove particular trend derived with stochastic geometry has been already used in \cite{GridModel}; in our case this verification has a significant additional value to the match with the trend in the experimental setting.

\section{Impact on architecture and system design} \label{architecture}

\begin{figure}[h!]
	\centering
		\includegraphics[width=16cm]{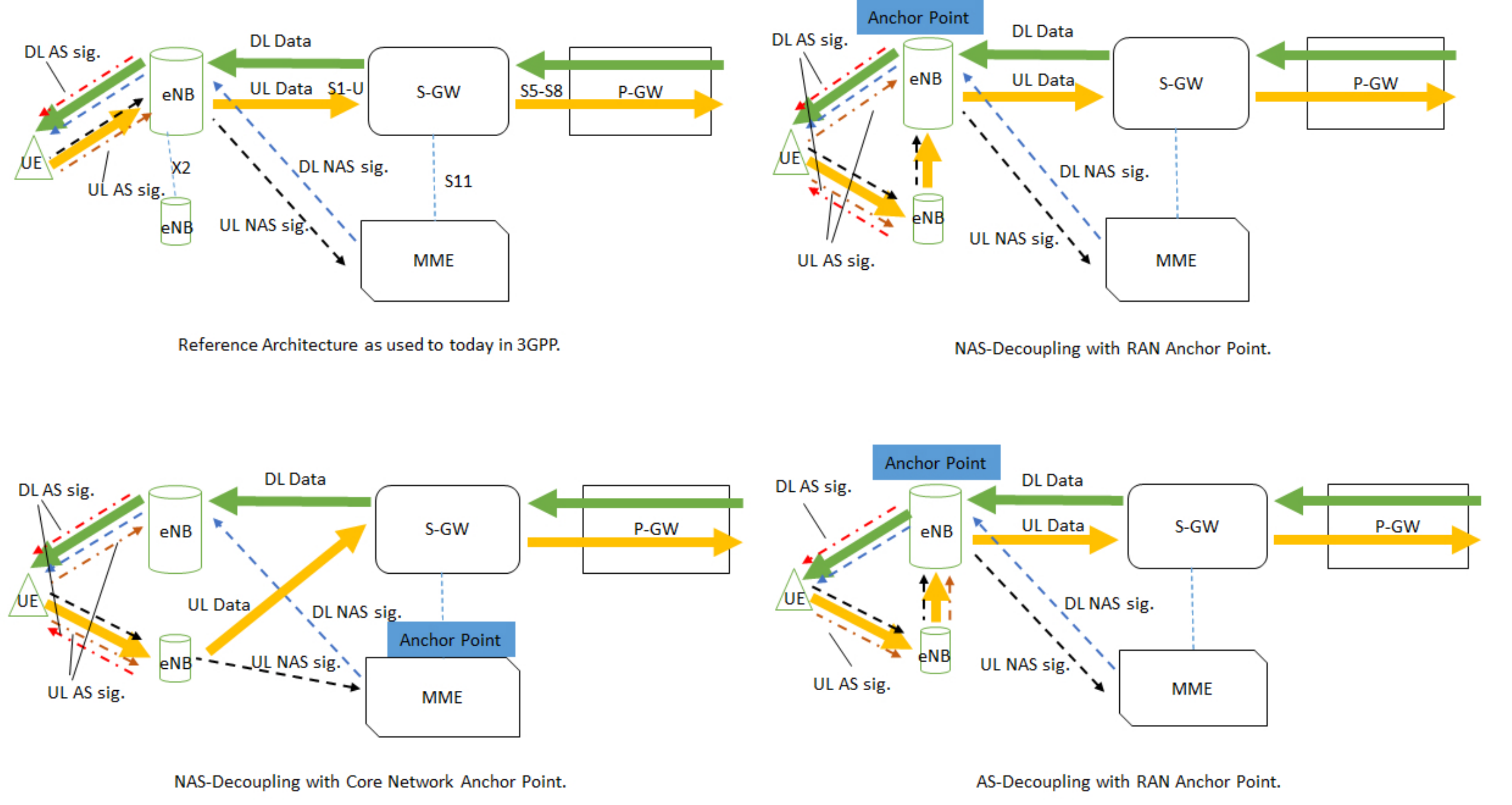}
		\caption{Suggested DUDe architecture embodiments.}
		\label{fig:DUDe_architecture}
\end{figure}

Decoupling the downlink from the uplink naturally requires some changes in the overall architecture, spanning from the radio access network (RAN) to the core network (CN). Decoupling flow notably requires some tight control to enable a smooth flow splitting and flow reassembly.
From a physical channel perspective, two possible solutions have been proposed in \cite{Ericsson_DUDe} during the Rel. 12 works. Herein we study DUDe-enabling architectures from the perspective of Access-Stratum (AS) and Non-Access Stratum (NAS) signaling. AS signaling refers to Layer 1, Layer 2 and RRC control messages exchanged between UE and BS. NAS signaling refers to control messages exchanged between UE and core-network.  It includes e.g. establishing and managing bearers, authentication and identification messages, mobility management and tracking area update \cite{Olsson_2009}. We note that delay, reliability and throughput requirements for each of these are very different, which naturally leads to a plethora of possible architectures.

To this end, we propose a few options which rely on the following assumptions:
\begin{itemize}
\item  a baseline 3GPP architecture;
\item availability of Multi-Flow TCP, which is able to handle different data flows at networking layer;
\item CoMP, which defines RAN traffic anchor points;
\item  support of multi-homing which allows different RATs to be connected at the same time to the same UE.
\end{itemize}
The options discussed below are mainly differentiated in the layer where the separation occurs, as well as the anchor point of choice where the traffic is reunited again. The different choices are depicted in Fig. \ref{fig:DUDe_architecture} and discussed below

\begin{itemize}
	\item
	{\bf NAS-Decoupling with RAN Anchor Point.} In this proposal a dedicated AS bi-directional connection is kept for both the Mcell and the Scell. The strength of this architecture is that all the delay-sensitive signaling (such as H-ARQ signaling) is handled by each cell. However, this requires the use of bidirectional physical control-channels. Moreover, we note that having the traffic-merging anchor point in the RAN requires the traffic to be handled via an established X2 interface, similar to the CoMP procedures currently outlined in 3GPP. NAS signaling is also handled via the X2 interface.
	\item
	{\bf NAS-Decoupling with CN Anchor Point.} This proposal differs from the previous one for the fact that data and NAS signaling are directly routed from the Scell to the core. The advantage here is that mobility is handled in a more efficient way, at the expense of delays due to the MME, which often resides physically far from the RAN. Also in this case, delay-sensitive signaling is handled via bidirectional exchanges at each cell.

	\item
	{\bf AS-Decoupling with RAN Anchor Point.} In this embodiment, the most aggressive of all, there is a complete separation of the traffic; i.e. if the UE communicates in the UL to the Scell, no DL is maintained. This requires AS and NAS information to be sent with minimal delay via the DL of the Mcell. The disadvantage here is that the X2 needs to facilitate close-to-zero delay communications; the advantage is that radio capacity is completely freed in the decoupled link.
\end{itemize}

We note that we didn't consider the case of AS-Decoupling with CN Anchor Point, due to the fact that for delay-sensitive control signaling is not possible to tolerate delays due to the anchor in the CN often physically residing far from the RAN.

A detailed analysis of the different alternatives is out-of-scope of this paper and it will be the object of our future work.

\section{Concluding Remarks} \label{conclusions}

The decoupling of the downlink and uplink is an emerging paradigm which will likely impact 5G design efforts. To this end, in this paper, we have rigorously characterized the capacity gains of a
system in which device associates to the infrastructure following the principle of Downlink and Uplink Decoupling (DUDe). The analysis was based on prior derivations on the association probabilities for the different uplink and downlink cases to the Macro or Small cells and it was extended towards calculating the transmission capacity. The analytical expressions have also been verified w.r.t. simulation results.

The work was then extended towards the evaluation of the decoupling framework within realistic real-world settings. To this end, we made use of a system level simulator Atoll from Vodafone. Said tool considers realistic traffic maps, realistic channel conditions, power control, etc. We have evaluated the association probability and absolute performance gains, in terms of throughput. Whilst absolute values differed slightly, it is important and surprising to observe that the trends of the real-world simulator coincided with the ones of the theoretical framework in the first part of the paper. This lead us to the following conjecture: the association probability depends chiefly on the density of the deployment, but not the process used to generate the deployment geometry. We have tested this conjecture with a third deployment setup, based on a regular grid, which corroborated the trend in the association probability.

In terms of performance improvements, a DUDe-enabled system yields gains of more than $200$\% and $600$\% for the $5$th percentile (cell edge) and $50$th percentile users, respectively. The
effects of the decoupled access is mainly pronounced in the $5$th and $50$th percentile of the experimental setting. Comparing Pcell and Fcell deployment, both parties show that Pcells, with their higher transmit power, introduce a fairness in the association process, attracting more devices to the Scell tier. The higher the disparity between the transmit power of Mcell and Fcell tier, the higher the need for DUDe and therefore the higher the gain from DUDe.

We have also outlined the architectural changes required to facilitate the decoupled UL/DL. Based on a 3GPP architecture, we proposed several different degrees of decoupling which mainly differ in the rigor of decoupling and the chosen anchor points in the RAN/core.

We believe that the outlined analysis only scratches the surface of the iceberg of possible RAN/core improvements in 5G systems. Combining the DUDe approach with other emerging paradigms, such as COMP, mm-Wave, D2D and Massive MIMO, would be interesting items for future work.

%
%
%

\end{document}